\documentclass[10pt,a4paper]{article}
\usepackage{jheppub_kim}
\usepackage{pdflscape}
\usepackage{amsmath}
\usepackage{amssymb}
\usepackage{dcolumn}
\usepackage{bm}
\usepackage{color}
\usepackage{epsfig}
\usepackage{amsfonts}
\usepackage{graphicx}
\usepackage{subfigure}
\usepackage{dcolumn}
\UseRawInputEncoding

\begin{document}
\title{Ricci-cubic holographic dark energy}

\author[a]{Prabir Rudra}

\affiliation[a] {Department of Mathematics, Asutosh College,
Kolkata-700 026, India}

\emailAdd{prudra.math@gmail.com, prabir.rudra@asutoshcollege.in}

\abstract{In this work, we propose the Ricci-cubic holographic
dark energy model. The model is inspired by the cubic curvature
invariant formed by the contraction of three Riemann tensors. A
combination of Ricci scalar and the cubic invariant is used to
describe the infrared cutoff of the holographic dark energy. Such
a construction is extremely useful since the evolution does not
depend on the past or future features of the universe, but
completely on its present features. Moreover, the use of
invariants makes the theory more fundamental in nature. We have
constructed the model and studied its cosmological features. The
analytical solutions of various cosmological parameters such as
the density parameter, equation of state parameter, and
deceleration parameter are extracted and their behaviour is
studied. It is seen that the holographic dark energy model can
exhibit all the cosmological epoch, sequentially starting from
radiation in the early universe, followed by matter, and finally
the dark energy dominated epoch at late times. The equation of
state parameter shows that the model can exhibit quintessence
nature, phantom-divide crossing, and even phantom nature depending
on the choice of parameter spaces.}


\maketitle

%%%%%%%%%%%%%%%%%%%%%%%%%%%%%%%%%%%%%%%%%%%%
\section{Introduction}
%%%%%%%%%%%%%%%%%%%%%%%%%%%%%%%%%%%%%%%%%%%
Observations from the SN Ia supernovae have confirmed that of
late, the universe has entered in an accelerated expansion phase
\cite{acc1, acc2}. It is obvious that the long standing matter
dominated epoch has come to an end. There are various theoretical
framework which can explain this unusual phenomenon, of which, the
most usual one is the cosmological constant $\Lambda$
\cite{cosmo1}. But in order to find a solution of the dynamical
nature one needs to introduce extra degrees of freedom beyond the
standard framework of general relativity (GR) \cite{gr} and
standard model of particle physics. Moreover explaining the entire
thermal history of the universe including the early time inflation
is always an issue. Modification of GR can be done via two
different avenues. The first one is by introducing modifications
in the geometrical sector giving rise to \textit{modified gravity
theories} \cite{mod1, mod2, mod3}. The other way is to modify the
matter sector thus introducing exotic components with negative
pressure known as \textit{dark energy} (DE) \cite{de1}. Both these
concepts employ extra degrees of freedom as desirable.

There are various candidates of DE available in the literature.
Chaplygin gas models \cite{cg1, cg2, cg3, cg4, cg5} and scalar
field models are notable examples. The holographic principle,
which has its origin in the black hole thermodynamics, states that
the entropy of a system is characterized by its area and not by
its volume \cite{holo1, holo2}. Holographic dark energy (HDE)
\cite{hde1, hde2, hde3} has been developed in connection with this
holographic principle, which also has connections with the string
theory \cite{holo1, string1}. It is known that a quantum field
theory has connections with an ultraviolet cutoff which is the
largest distance possible under the framework \cite{uv1}. This
ultraviolet cutoff in turn has direct connection with the vacuum
energy, which will be a form of dark energy of the holographic
origin. For an extensive review on HDE the reader may refer to
\cite{hderev1}. There have been extensive research on HDE both in
its basic and extended forms and with time the model has been
quite successful \cite{hde4, hde5, hde6, hde7, hde8, hde9, hde10}.
One of the major success of HDE models have been its compatibility
with the observational data \cite{data1, data2, data3}.

It is an accepted fact that the HDE density is proportional to the
inverse square of the infrared cutoff $L$ given by,
\begin{equation}\label{hdedensity}
\rho_{DE}=\frac{3c}{\kappa^{2}L^{2}}
\end{equation}
where $\kappa^{2}$ is the gravitational constant and $c$ is a
parameter. However in connection with the cosmological application
of the holographic principle there is no accepted idea about what
the infrared cutoff should be. The most common choices are the
Hubble radius and the particle horizon which are incapable of
driving the cosmic acceleration \cite{hubh}. Finally it is the
future event horizon that suits the scenario and can suitably act
as the infrared cutoff \cite{hde3}. Although this choice suits the
scenario well, there are some logical problems associated with it.
The present value of the dark energy density is actually
determined by the future evolution of the DE, which is quite an
uncomfortable concept to deal with. So further attempts have been
made in the quest of finding modified holographic dark energy
models, where the DE density does not depend on the future
evolution, but on the past and present evolution. Models where the
infrared cutoff can be given by the quantities depending on the
past features of the universe are called the \textit{agegraphic
dark energy} and here the infrared cutoff is given by the age of
the universe or the conformal time \cite{age1, age2, age3}. The
model which depend upon the present evolution of the universe
involves the use of the inverse square root of Ricci scalar as the
infrared cutoff \cite{age4}. This is called the Ricci holographic
dark energy. Using Ricci scalar as the infrared cutoff has the
added advantage that the evolution of the dark energy is governed
by a gravitational invariant which has fundamental importance in
gravitational theories. Moreover its evolution depends only on the
present features of the universe, which is theoretically much more
sound concept. Apart from this, Ricci holographic dark energy has
very interesting cosmological applications \cite{age4, ricc1,
ricc2, ricc3}. Some generalized features of holographic dark
energy can be found in \cite{od1, od2, od3, od4, od5}. Some
phenomenological problems of minimal holographic dark energy are
given in \cite{od6, od7}.

Since the use of invariants is fundamental in physics, naturally a
we can use other invariants to represent the infrared cutoff and
thus give rise to new forms of holographic dark energy. The
simplest such extension would be using the Gauss-Bonnet invariant.
Saridakis in \cite{rgbhde} addressed this problem by using a
combination of the Ricci scalar and Gauss-Bonnet invariant to
describe the infrared cutoff as given below,
\begin{equation}
\frac{1}{L^{2}}=-\alpha R+\beta \sqrt{|G|}
\end{equation}
This model of HDE was called the Ricci-Gauss Bonnet HDE, and it
exhibited very interesting cosmological features including the
entire thermal history of the universe, from the radiation
followed by matter, and finally dark energy dominated epoch.

Since we are talking about invariants, there can be many such
possibilities from the mathematical point of view. The
Gauss-Bonnet invariant was formed from the contractions of two
Riemann tensors. So it is a quadratic theory whose spectrum
coincides with that of the Einstein gravity. We can always play
around with the Riemann tensor $R_{\mu\nu\rho\sigma}$, Ricci
tensor $R_{\mu\nu}$, energy momentum tensor $T_{\mu\nu}$, etc. and
explore their possible contractions and produce different scalar
invariants from such exercises. It is understandable that their
physical significance and importance to cosmology is a different
issue and needs thorough study, their mathematical importance is
irrefutable. For the cubic case there exists a six parameter
family of cubic theories whose spectrum is identical to that of
the Einstein gravity \cite{cub11}. In 4-dimensional spacetime a
general non-topological cubic term would be given by \cite{cub11,
cub1}

$$P=\beta_{1}R_{\mu~~\nu}^{~~\rho~~\sigma}R_{\rho~~\sigma}^{~~\gamma~~\delta}R_{\gamma~~\delta}^{~~\mu~~\nu}
+\beta_{2}R_{\mu\nu}^{\rho\sigma}~R_{\rho\sigma}^{\gamma\delta}~R_{\gamma\delta}^{\mu\nu}+\beta_{3}R^{\sigma\gamma}R_{\mu\nu\rho\sigma}{R^{\mu\nu\rho}}_{\gamma}
+\beta_{4}RR_{\mu\nu\rho\sigma}R^{\mu\nu\rho\sigma}$$
\begin{equation}\label{cubicterm}
+\beta_{5}R_{\mu\nu\rho\sigma}R^{\mu\rho}R^{\nu\sigma}+\beta_{6}R^{\nu}_{\mu}R^{\rho}_{\nu}R^{\mu}_{\rho}+\beta_{7}R_{\mu\nu}R^{\mu\nu}R+\beta_{8}R^{3}
\end{equation}
where $\beta_{i}$ are parameters. If the theory possesses a
spectrum identical to that of general relativity then the
following parameter conditions are satisfied,
\begin{equation}\label{cond1}
\beta_{7}=\frac{1}{12}\left(3\beta_{1}-24\beta_{2}-16\beta_{3}-48\beta_{4}-5\beta_{5}-9\beta_{6}\right)
\end{equation}
\begin{equation}\label{cond2}
\beta_{8}=\frac{1}{72}\left(-6\beta_{1}+36\beta_{2}+22\beta_{3}+64\beta_{4}+3\beta_{5}+9\beta_{6}\right)
\end{equation}
The cubic invariant has been used to develop the cubic gravity
theory \cite{cub11} and even extended to generalized $f(P)$
gravity theories \cite{cub1, cub2, cub3}. Motivated from
\cite{rgbhde}, here we are interested in constructing a
holographic dark energy model where the infrared cutoff is given
by a combination of Ricci scalar and the cubic invariant given in
eqn.(\ref{cubicterm}). This will be an extension of the Ricci
holographic dark energy and the Ricci Gauss-Bonnet holographic
dark energy. We term it as Ricci-cubic holographic dark energy and
here we are interested in constructing the model and exploring its
cosmological features. Such a construction is more generalized and
theoretically more tangible since higher order invariants have
their contributions in the set up. This will allow for a far
richer cosmological structure which is the basic motivation behind
this study. The paper is organized as follows: In section II we
construct the Ricci-cubic holographic dark energy and present the
basic equations. Section III is dedicated to the cosmological
evolution of the universe filled with the HDE model. Finally the
paper ends with a discussion and conclusion in section IV.

%%%%%%%%%%%%%%%%%%%%%%%%%%%%%%%%%%%%%%%%%
\section{Ricci-cubic holographic dark energy}
%%%%%%%%%%%%%%%%%%%%%%%%%%%%%%%%%%%%%%%%%
Here we will construct a model of holographic dark energy where
the infrared cutoff is given by a combination of the Ricci and
cubic scales. We will consider a homogeneous and isotropic
universe modelled by the Friedmann-Lemaitre-Robertson-Walker
(FLRW) metric given by,
\begin{equation}\label{flrwmet}
ds^{2}=-dt^{2}+a^{2}(t)\left(\frac{dr^{2}}{1-kr^{2}}+r^{2}d\Omega^{2}\right)
\end{equation}
where $a(t)$ is the cosmological scale factor and $k$ is the
spatial curvature, such that $k=-1, 0, +1$ corresponds to open,
flat and closed spatial geometry respectively.
$d\Omega^{2}=d\theta^{2}+\sin^{2}\theta d\phi^{2}$ represents the
2-sphere. Here we will concentrate on the flat geometry which may
be easily extended to the open and closed universes.

It is known that Ricci holographic dark energy uses the Ricci
scalar $R$ in the FLRW metric as the infrared cutoff. Since the
dimensions of the Ricci scalar is $1/(Length)^{2}$, the energy
density of the dark energy is obtained proportional to $R$.
Whenever we use a modification using any curvature invariants, it
is imperative to use the new invariants in the same order in order
to preserve the consistency. We know that in the FLRW geometry the
cubic invariant $P$ occurs in the order of $R^{3}$. Therefore it
can be argued that in any model of HDE where the IR cutoff is
given by $R$, there should be contributions from $P^{1/3}$ also.
Following this motivation here we will consider an HDE model where
the inverse square of the IR cutoff is given by,
\begin{equation}\label{irrc}
\frac{1}{L^{2}}=-\alpha R+\lambda P^{1/3}
\end{equation}
where the constants $\alpha$ and $\lambda$ are model parameters.
It can be clearly seen that for $\lambda=0$, we retrieve the usual
Ricci HDE, while for $\alpha=0$, we get a pure cubic HDE. Using
eqn.(\ref{irrc}) in eqn.(\ref{hdedensity}) we get the energy
density of the Ricci-cubic HDE as,
\begin{equation}\label{denrc1}
\rho_{DE}=\frac{3}{\kappa^{2}}\left(-\alpha R+\lambda
P^{1/3}\right)
\end{equation}
where the constant $c$ has been absorbed in the model parameters
$\alpha$ and $\lambda$ for convenience. Now for the flat FLRW
geometry the Ricci scalar and the cubic invariant are respectively
given by,
\begin{equation}\label{ricfrw}
R=-6\left(2H^{2}+\dot{H}\right)
\end{equation}
and
\begin{equation}\label{cubfrw}
P=6\tilde{\beta}H^{4}\left(2H^{2}+3\dot{H}\right)
\end{equation}
where $H=\dot{a}/a$ is the Hubble function and the dots denote
derivatives with respect to time. Moreover in the above expression
we have defined $\tilde{\beta}$ as,
\begin{equation}
\tilde{\beta}\equiv -\beta_{1}+4\beta_{2}+2\beta_{3}+8\beta_{4}
\end{equation}
It should be mentioned here that for the cubic invariant we have
considered derivatives only upto first order, so that the FLRW
equations are of the second order. The condition for the second
order field equations is satisfied if we consider,
\begin{equation}
\beta_{6}=4\beta_{2}+2\beta_{3}+8\beta_{4}+\beta_{5}
\end{equation}
Using these, the energy density of Ricci-cubic HDE becomes,
\begin{equation}\label{denrc2}
\rho_{DE}=\frac{3}{\kappa^{2}}\left[6\alpha\left(2H^{2}+\dot{H}\right)+\lambda
\left\{6\tilde{\beta}H^{4}\left(2H^{2}+3\dot{H}\right)\right\}^{1/3}\right]
\end{equation}
Now the first FLRW equation is given by,
\begin{equation}\label{flrweq1}
3H^{2}=\kappa^{2}\left(\rho_{m}+\rho_{DE}\right)
\end{equation}
where $\rho_{m}$ is the energy density of matter. The equation of
state (EoS) parameter of matter is given by
$w_{m}=p_{m}/\rho_{m}$, where $p_{m}$ is the pressure of matter.
Finally the matter sector follows the conservation relation given
by,
\begin{equation}\label{conseq}
\dot{\rho}_{m}+3H\left(\rho_{m}+p_{m}\right)=0
\end{equation}
Using eqns.(\ref{flrweq1}) and (\ref{conseq}) one can fully
determine the evolution of the universe, provided the matter
equation of state is known. Generally we consider a pressureless
matter sector, i.e. $p_{m}=0$ leading to $w_{m}=0$. Using this in
the conservation equation (\ref{conseq}) we get the matter energy
density as $\rho_{m}=\rho_{m0}a^{-3}$, where $\rho_{m0}$ is the
density of matter in the present time. Thus we have the matter
energy density and also from eqn.(\ref{denrc2}) we have the dark
energy density. Using them in the FLRW equation (\ref{flrweq1}) we
have a differential equation in terms of the scale factor $a$,
which may be solved to get the evolution of the universe filled
with Ricci-cubic holographic dark energy. In the following section
we will study the cosmological evolution of such a universe.

%%%%%%%%%%%%%%%%%%%%%%%%%%%%%%%%%
\section{Cosmological Evolution}
%%%%%%%%%%%%%%%%%%%%%%%%%%%%%%%%%
In this section we will study the cosmological evolution of a
universe filled with Ricci-cubic HDE and pressureless matter in
the form of dust. Let us introduce the density parameters,
\begin{equation}\label{denpa}
\Omega_{m}\equiv
\frac{\kappa^{2}\rho_{m}}{3H^{2}},~~~~~~~~~~~\Omega_{DE}\equiv
\frac{\kappa^{2}\rho_{DE}}{3H^{2}}
\end{equation}
In terms of the above density parameters the FLRW equation
(\ref{flrweq1}) becomes $\Omega_{m}+\Omega_{DE}=1$. Using the
energy density for dust we can write $\Omega_{m}$ as
$\Omega_{m}=\Omega_{m0}H_{0}^{2}/H^{2}a^{3}$, where $\Omega_{m0}$
is the present value of $\Omega_{m}$ given by
$\Omega_{m0}=\kappa^{2}\rho_{m0}/3H_{0}^{2}$. Similarly $H_{0}$ is
the present value of the Hubble function. Using this result we can
easily get the Hubble function as below,
\begin{equation}\label{hub}
H=\frac{H_{0}\sqrt{\Omega_{m0}}}{\sqrt{a^{3}\left(1-\Omega_{DE}\right)}}
\end{equation}
Here we will use $x=\ln a$ as the independent variable.
Differentiating eqn.(\ref{hub}) we have,
\begin{equation}\label{hdot}
\dot{H}=-\frac{H^{2}}{2\left(1-\Omega_{DE}\right)}\left[3\left(1-\Omega_{DE}\right)-\Omega_{DE}'\right]
\end{equation}
where prime represents derivative with respect to $x$. Using the
above equation in eqns.(\ref{ricfrw}) and (\ref{cubfrw}) we get,
\begin{equation}
R=-3H^{2}\left(1+\frac{\Omega_{DE}'}{1-\Omega_{DE}}\right)
\end{equation}
\begin{equation}
P=3\tilde{\beta}H^{6}\left(\frac{3\Omega_{DE}'}{1-\Omega_{DE}}-5\right)
\end{equation}
Using the above expressions for $R$ and $P$ in eqn.(\ref{denrc1})
we get,
\begin{equation}
\rho_{DE}=\frac{3H^{2}}{\kappa^2}\left[3^{1/3}\lambda
\left(\frac{\tilde{\beta}\left(5-5\Omega_{DE}-3\Omega_{DE}'\right)}{\Omega_{DE}-1}\right)^{1/3}+\frac{3\alpha\left(-1+\Omega_{DE}-\Omega_{DE}'\right)}{\Omega_{DE}-1}\right]
\end{equation}
Putting the above expression for energy density in (\ref{denpa})
we get the dimensionless energy density as,
\begin{equation}\label{energydiff}
\Omega_{DE}-3^{1/3}\lambda
\left(\frac{\tilde{\beta}\left(5-5\Omega_{DE}-3\Omega_{DE}'\right)}{\Omega_{DE}-1}\right)^{1/3}-\frac{3\alpha\left(-1+\Omega_{DE}-\Omega_{DE}'\right)}{\Omega_{DE}-1}=0
\end{equation}
The above differential equation governs the evolution of the
Ricci-cubic HDE for a flat universe with matter in the form of
dust. Unfortunately this equation does not have any general
analytic solution. So we try to find out a solution under some
assumptions. The first term on the right hand side being a cube
root over $\Omega_{DE}$ and its derivatives is the most
complicated term. So we expand the term binomially and consider
only the linear powers of $\Omega_{DE}$ and its derivative. We are
quite justified in doing this since it can be considered that the
energy density of any component of the universe in the late
universe should be low enough, so that higher powers may be
neglected. A solution for the simplified equation is obtained as,
\begin{equation}\label{desol1}
\Omega_{DE}=\frac{\left(3\alpha-\xi\right)e^{3\alpha\left(\frac{5\ln{a}}{15\alpha+\xi}+C_{1}\right)}-e^{\frac{\left(1+\xi\right)\left(5\ln{a}+C_{1}\left(15\alpha+\xi\right)\right)}{15\alpha+\xi}}}{e^{3\alpha\left(\frac{5\ln{a}}{15\alpha+\xi}+C_{1}\right)}-e^{\frac{\left(1+\xi\right)\left(5\ln{a}+C_{1}\left(15\alpha+\xi\right)\right)}{15\alpha+\xi}}}
\end{equation}
where $\xi=(15\tilde{\beta})^{1/3}\lambda$ and $C_{1}$ is the
constant of integration. The integration constant can easily
determined by considering the scale factor of the present universe
as $a=a_{0}=1$ and the corresponding energy density as
$\Omega_{DE}=\Omega_{DE0}$. So eqn.(\ref{desol1}) gives the
evolution of the Ricci-cubic HDE in terms of the logarithm of the
scale factor. This can easily be evaluated in terms of the
redshift by using the relation $z=\frac{a_{0}}{a-1}$, since
$x\equiv \ln{a}=-\ln{(1+z)}$. Finally using eqn.(\ref{desol1}) in
eqn.(\ref{hub}) we can easily obtain the Hubble function, which
can further be integrated to obtain the scale factor $a(t)$.

Another important parameter regarding the dark energy model is the
equation of state parameter given by $w_{DE}=p_{DE}/\rho_{DE}$,
where $p_{DE}$ and $\rho_{DE}$ are the pressure and density of
holographic dark energy respectively. Since matter is conserved
according to eqn.(\ref{conseq}), Ricci-cubic HDE will be conserved
according to the equation,
\begin{equation}\label{conseqhde}
\dot{\rho}_{DE}+3H\left(1+w_{DE}\right)\rho_{DE}=0
\end{equation}
Using eqn.(\ref{denrc1}) in the above equation we get
\begin{equation}\label{omegadef}
w_{DE}=-1+\Omega_{DE}^{-1}\left(\frac{\alpha
R'}{3H^2}-\frac{\lambda P'}{9H^{2}P^{2/3}}\right)
\end{equation}
Now differentiating eqns.(\ref{ricfrw}) and (\ref{cubfrw}) and
using eqn.(\ref{hdot}) we have,
\begin{equation}\label{rdashf}
\frac{R'}{3H^2}=3+\frac{2\Omega_{DE}'}{1-\Omega_{DE}}-\frac{2(\Omega_{DE}')^2}{\left(1-\Omega_{DE}\right)^2}-\frac{\Omega_{DE}''}{1-\Omega_{DE}}
\end{equation}
and
\begin{equation}\label{pdashf}
\frac{P'}{9H^{2}P^{2/3}}=\frac{\tilde{\beta}^{1/3}}{3^{2/3}\left(1-\Omega_{DE}\right)^2}\frac{\left[\Omega_{DE}''\left(1-\Omega_{DE}\right)
+15\left(1-\Omega_{DE}\right)^{2}+2\Omega_{DE}'\left(2\Omega_{DE}'+7\Omega_{DE}-7\right)\right]}{\left(\frac{3\Omega_{DE}'}{1-\Omega_{DE}}-5\right)^{2/3}}
\end{equation}
Using eqns.(\ref{rdashf}) and (\ref{pdashf}) in
eqn.(\ref{omegadef}) we get the EoS parameter as,

$$w_{DE}=-1-\Omega_{DE}^{-1}\left[\alpha\left\{3+\frac{2\Omega_{DE}'}{1-\Omega_{DE}}-\frac{2(\Omega_{DE}')^2}{\left(1-\Omega_{DE}\right)^2}
-\frac{\Omega_{DE}''}{1-\Omega_{DE}}\right\}\right.$$

\begin{equation}\label{eosrgbhde}
\left.-\frac{\lambda\tilde{\beta}^{1/3}}{3^{2/3}\left(1-\Omega_{DE}\right)^2}\frac{\left\{\Omega_{DE}''\left(1-\Omega_{DE}\right)
+15\left(1-\Omega_{DE}\right)^{2}+2\Omega_{DE}'\left(2\Omega_{DE}'+7\Omega_{DE}-7\right)\right\}}{\left(\frac{3\Omega_{DE}'}{1-\Omega_{DE}}-5\right)^{2/3}}\right]
\end{equation}
This expression gives the EoS parameter of the Ricci-cubic HDE in
terms of $\ln{a}$, i.e. as a function of the redshift. Finally we
can introduce the deceleration parameter which is crucial for any
model of dark energy, and is given by,
\begin{equation}
q=-1-\frac{\dot{H}}{H^2}=\frac{1}{2}+\frac{3}{2}\left(w_{m}\Omega_{m}+w_{DE}\Omega_{DE}\right)
\end{equation}

In Fig.(1) we present the dimensionless density parameters
$\Omega_{DE}(z)$ and $\Omega_{m}(z)=1-\Omega_{DE}(z)$ as functions
of the redshift $z$. We see that as the universe evolves the dark
energy dominates over the matter sector, which complies with known
results. In Fig.(2) we have plotted the equation of state (EoS)
parameter against the redshift as it arises from
eqn.(\ref{eosrgbhde}). From the plot wee see that presently
($z=0$) the universe has entered a dark energy dominated phase
($w_{DE}<-1/3$). In fact it has the plunged deep into the DE
dominated region with a fair possibility of phantom-divide
crossing ($w_{DE}<-1$). In Fig.(3) the deceleration parameter $q$
have been plotted against the redshift $z$. It is clearly seen
that the usual evolution of the universe can be obtained from the
Ricci-cubic holographic dark energy with the transition from
deceleration to acceleration occurring at $z\approx 0.45$ as the
observations suggest.

From eqn.(\ref{omegadef}) it is seen that even if
$\Omega_{DE}\rightarrow 1$ in the future universe ($z<0$), the
asymptotic value of $w_{DE}$ still depends on the model parameters
$\alpha$ and $\lambda$. So the effects of these parameters on the
EoS parameter is significant, and it must be studied in detail. So
we have plotted $w_{DE}$ for various choices of $\lambda$ and
$\alpha$ in figs.(4) and (5) respectively. In Fig.(4) we have
presented the EoS parameter against the redshift for different
values of the parameter $\lambda$, which scales the cubic term $P$
in the energy density of the Ricci-cubic HDE. From the plot we see
that with the increase in the value of $\lambda$ there is a
decrease in both the current ($z=0$) and future ($z<0$) values of
the EoS parameter. Moreover it is seen that the dependence of the
EoS parameter on $\lambda$ is highly pronounced in the future
epoch and declines both in the present and past times. It is
evident from the figure that in all the trajectories there is a
transition from $w_{DE}>-1$ to $w_{DE}<-1$ around $z\approx 0.45$.
This means that the universe has evolved from the quintessence
regime to the phantom regime of late (around the present time). We
can also comprehend that for proper parameter spaces, we can have
trajectories lying completely in the quintessence regime going
asymptotically to $w_{DE}\rightarrow -1$, resulting in a de-Sitter
universe. This is because from the eqns.(\ref{denrc2}) and
(\ref{flrweq1}) we see that a late de-Sitter like evolution is
possible if
$12\alpha+\left(12\tilde{\beta}\right)^{1/3}\lambda=1$. Similarly
for suitable parameter spaces, it is also possible to obtain
trajectories lying completely in the phantom regime ($w_{DE}<-1$),
where the universe is driven towards a Big Rip. In Fig.(5) we have
plotted the EoS parameter against the redshift for different
values of the parameter $\alpha$, which scales the Ricci scalar
component of the HDE. It is see that with an increase in the value
of $\alpha$ there is a corresponding decrease in the EoS
parameter. We also see that the evolution of the EoS parameter
depends significantly on $\alpha$ in the future epoch ($z<0$). It
should be mentioned that in all these plots we have fixed the
constant $C_{1}$ in order to obtain
$\Omega_{DE}(z=0)=\Omega_{DE0}\approx 0.68$ consistent with the
observations \cite{obs1}. To gain more insights into the future
evolution of the universe described by Ricci-cubic HDE and to
understand the stability of the various asymptotic solutions one
must employ the dynamical system analysis. Such a study is beyond
the scope of the present work and can be a potential future
project.

\begin{figure}\label{g10}
~~~~~~~~~~~~~~~~~~~~~~~~~~~~~~\includegraphics[height=2.2in,width=3.5in]{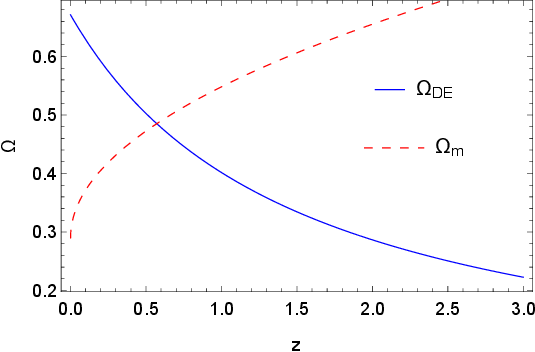}~~~~~\\

~~~~~~~~~~~~~~~~~~~~~~~~~~~~~~~~~~~~~~~~~~~~~~~~~~~~~~~~~~~~Fig.1~~~~~~~~~~~~~~~~~~~~\\

\vspace{1mm} \textit{\textbf{Fig.1} shows the evolution of the
Ricci-cubic holographic dark energy density parameters
$\Omega_{DE}$ and $\Omega_{m}$ as a function of the redshift $z$
for $\alpha=-0.45$, $\lambda=1.09$ and $\tilde{\beta}=100$. The
constant of integration $C_{1}$ has been fixed in order to obtain
$\Omega_{DE}(z=0)\equiv\Omega_{DE0}\approx 0.68$ at present time.}
\end{figure}

\begin{figure}\label{fig2}
~~~~~~~~~~~~~~~~~~~~~~~~~~~~~~\includegraphics[height=2.2in,width=3.5in]{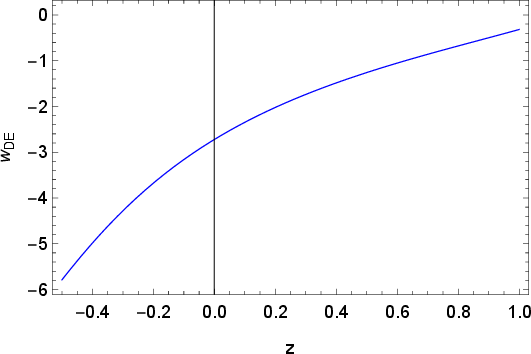}~~~~~\\

~~~~~~~~~~~~~~~~~~~~~~~~~~~~~~~~~~~~~~~~~~~~~~~~~~~~~~~~~~~~Fig.2~~~~~~~~~~~~~~~~~~~~\\

\vspace{1mm} \textit{\textbf{Fig.2} shows the evolution of the
equation of state (EoS) parameter $w_{DE}$ of Ricci-cubic
holographic dark energy as a function of the redshift $z$ for
$\alpha=1.5$, $\lambda=0.09$, $\tilde{\beta}=10^6$. The constant
of integration $C_{1}$ has been fixed in order to obtain
$\Omega_{DE}(z=0)=\Omega_{DE0}\approx 0.68$.}
\end{figure}

\begin{figure}\label{fig3}
~~~~~~~~~~~~~~~~~~~~~~~~~~~~~~\includegraphics[height=2.2in,width=3.5in]{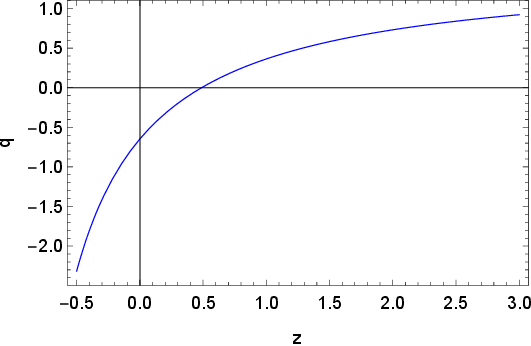}~~~~~\\

~~~~~~~~~~~~~~~~~~~~~~~~~~~~~~~~~~~~~~~~~~~~~~~~~~~~~~~~~~~~Fig.3~~~~~~~~~~~~~~~~~~~~\\

\vspace{1mm} \textit{\textbf{Fig.3} shows the evolution of the
deceleration parameter $q$ as a function of the redshift $z$ for
$\alpha=0.1$, $\lambda=0.5$ and $\tilde{\beta}=100$. The constant
of integration $C_{1}$ has been fixed in order to obtain $q=0$ at
around $z\approx 0.45$ when the transition occurs.}
\end{figure}

If we consider $\lambda=0$, we get the usual Ricci dark energy and
the corresponding equations obtained above are considerably
simplified. From eqn.(\ref{energydiff}) we get $\Omega_{DE}$ in
the form,
\begin{equation}
\Omega_{DE|Ric}=\frac{e^{\frac{x}{3\alpha}+C_{2}}-3\alpha
e^{x+3\alpha C_{2}}}{e^{\frac{x}{3\alpha}+C_{2}}-e^{x+3\alpha
C_{2}}}
\end{equation}
which is exactly the form obtained in \cite{rgbhde}. In the above
expression $C_{2}$ is the constant of integration. Using this the
Hubble parameter in eqn.(\ref{hub}) becomes,
\begin{equation}
H_{Ric}=H_{0}\sqrt{\Omega_{m0}}e^{-\frac{3x}{2}}\left[\frac{\left(3\alpha-1\right)e^{x+3\alpha
C_{2}}}{e^{\frac{x}{3\alpha}+C_{2}}-e^{x+3\alpha
C_{2}}}\right]^{-1/2}
\end{equation}
Using the above expressions in eqn.(\ref{denpa}) we get,
\begin{equation}
\rho_{DE|Ric}=\frac{3H_{0}^{2}\Omega_{m0}}{\kappa^{2}\left(3\alpha-1\right)}
\left[e^{\left(\frac{1}{3\alpha}-4\right)x-C_{2}\left(3\alpha-1\right)}-3\alpha
e^{-3x}\right]
\end{equation}
This result exactly coincides with the expression given in
\cite{age4} with the constants being redefined. Finally the using
eqn.(\ref{eosrgbhde}) the EoS parameter for Ricci dark energy can
be given by,
\begin{equation}
w_{DE|Ric}=\frac{1}{3\Omega_{DE|Ric}}-\frac{1}{9\alpha}=\frac{3\alpha-1}{9\alpha\left(1-3\alpha
e^{\frac{\left(3\alpha-1\right)\left(x+3\alpha
C_{2}\right)}{3\alpha}}\right)}
\end{equation}
In the same way we can obtain pure cubic HDE by putting
$\alpha=0$. From eqn.(\ref{energydiff}) we get $\Omega_{DE}$ for
pure cubic HDE as,
\begin{equation}
\Omega_{DE|cub}=\frac{e^{5x+\zeta}+(15\beta)^{1/3}\lambda~
e^{C_{3}+(15\beta)^{1/3}\lambda
C_{3}}}{e^{5x+\zeta}-e^{C_{3}+(15\beta)^{1/3}\lambda C_{3}}}
\end{equation}
where $\zeta=\frac{5^{2/3}x}{(3\beta)^{1/3}\lambda}$ and $C_{3}$
is the constant of integration. The Hubble parameter in
eqn.(\ref{hub}) can be written as,
\begin{equation}
H_{cub}=H_{0}\sqrt{\Omega_{m0}}\left[\frac{e^{3x}e^{C_{3}+(15\beta)^{1/3}C_{3}\lambda}
\left(1+(15\beta)^{1/3}\lambda\right)}{e^{C_{3}+(15\beta)^{1/3}C_{3}\lambda}
-e^{\frac{1}{3}x\left(15+\frac{15^{2/3}}{\beta^{1/3}\lambda}\right)}}\right]^{-1/2}
\end{equation}
Using eqn.(\ref{denpa}) we get the energy density of cubic HDE as,
\begin{equation}
\rho_{DE|cub}=\frac{3H_{0}^{2}\Omega_{m0}e^{-3x}\left[e^{\frac{1}{3}x\left(15+
\frac{15^{2/3}}{\tilde{\beta}^{1/3}\lambda}\right)
-C_{3}\left(1+(15\tilde{\beta})^{1/3}\lambda\right)}+(15\tilde{\beta})^{1/3}\lambda\right]}
{\kappa^{2}\left(1+(15\tilde{\beta})^{1/3}\lambda\right)}
\end{equation}
The EoS parameter of pure cubic HDE can be obtained using
eqns.(\ref{desol1}) and (\ref{eosrgbhde}) as follows,
\begin{equation}
w_{DE|cub}=\frac{\Omega_{DE|cub}^{4}+15\Omega_{DE|cub}~\tilde{\beta}\lambda^{3}-27\Omega_{DE|cub}^{2}~
\tilde{\beta}\lambda^{3}-27\tilde{\beta}^{5/3}\lambda^{5}
\left(\frac{\Omega_{DE|cub}^3}{\tilde{\beta}\lambda^{3}}\right)^{2/3}}
{27\tilde{\beta}^{5/3}\lambda^{5}\left(\frac{\Omega_{DE|cub}^{3}}{\tilde{\beta}\lambda^{3}}\right)^{2/3}}
\end{equation}

Finally we will study two different cosmological set up using the
Ricci-cubic holographic dark energy. In the first case we will
explore an interacting scenario between the Ricci-cubic
holographic dark energy and dark matter. In the second case we
will include radiation sector in our set-up to describe the entire
thermal history of the universe.

\begin{figure}\label{fig4}
~~~~~~~~~~~~~~~~~~~~~~~~~~~~~~\includegraphics[height=2.2in,width=3.5in]{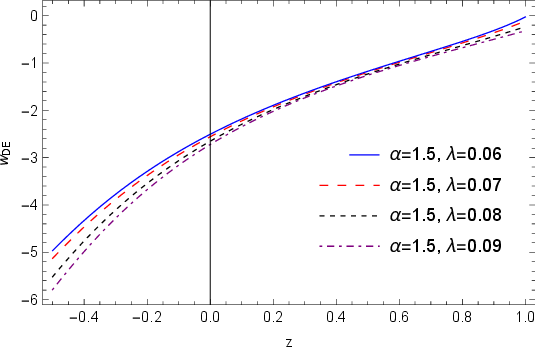}~~~~~\\

~~~~~~~~~~~~~~~~~~~~~~~~~~~~~~~~~~~~~~~~~~~~~~~~~~~~~~~~~~~~Fig.4~~~~~~~~~~~~~~~~~~~~\\

\vspace{1mm} \textit{\textbf{Fig.4} shows the evolution of the
equation of state (EoS) parameter $w_{DE}$ of Ricci-cubic
holographic dark energy as a function of the redshift $z$ for
different values of $\lambda$ keeping $\alpha$ unaltered. The
constant of integration $C_{1}$ has been fixed in order to obtain
$\Omega_{DE}(z=0)=\Omega_{DE0}\approx 0.68$. We have taken
$\tilde{\beta}=10^6$.}
\end{figure}

\begin{figure}\label{fig5}
~~~~~~~~~~~~~~~~~~~~~~~~~~~~~~\includegraphics[height=2.2in,width=3.5in]{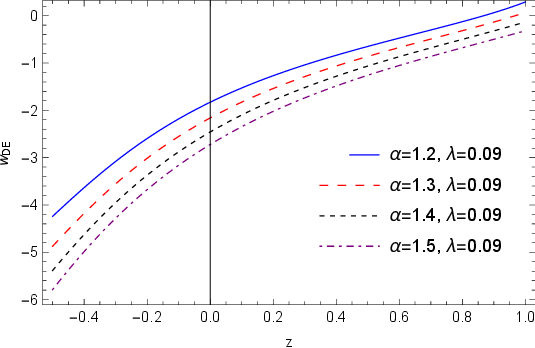}~~~~~\\

~~~~~~~~~~~~~~~~~~~~~~~~~~~~~~~~~~~~~~~~~~~~~~~~~~~~~~~~~~~~Fig.5~~~~~~~~~~~~~~~~~~~~\\

\vspace{1mm} \textit{\textbf{Fig.5} shows the evolution of the
equation of state (EoS) parameter $w_{DE}$ of Ricci-cubic
holographic dark energy as a function of the redshift $z$ for
different values of $\alpha$ keeping $\lambda$ unaltered. The
constant of integration $C_{1}$ has been fixed in order to obtain
$\Omega_{DE}(z=0)=\Omega_{DE0}\approx 0.68$. We have taken
$\tilde{\beta}=10^6$.}
\end{figure}

\subsection{Interacting scenario between Ricci-cubic HDE and dark matter}
A generalized scenario should allow for an interaction between
dark energy and dark matter sectors since it cannot be eliminated
by any logical arguments. Moreover inclusion of interaction in
cosmological models helps to alleviate the cosmic coincidence
problem, which raises question on the equality of the density
parameters of the dark energy and dark matter sectors although
these components have evolved following completely different
scales \cite{ccp1, ccp2}. Including the interacting term the
conservation equations for matter and HDE become respectively,
\begin{equation}\label{intm}
\dot{\rho}_{m}+3H\left(\rho_{m}+p_{m}\right)=-Q
\end{equation}
and
\begin{equation}\label{intde}
\dot{\rho}_{DE}+3H\left(1+w_{DE}\right)\rho_{DE}=Q
\end{equation}
where $Q$ represents the interaction between dark energy and dark
matter sectors. Here $Q>0$ indicates that there is a transfer from
the dark matter sector to the dark energy sector and $Q<0$
indicates the flow in the reverse direction. So a positive
interaction term basically means a growth the dark energy sector,
whereas a negative interaction indicates a decay of dark energy. A
satisfactory form of the interaction term is theorized in various
ways in the literature. Almost in all such forms it is evident
that the interaction should be proportional to the energy density
of the constituents of the universe. A well motivated form of
interaction widely studied in literature is $Q=\eta H\rho_{m}$,
where $\eta$ represents the rate of interaction between the
sectors \cite{ccp1, ccp2}. Now we proceed to explore the cosmology
of Ricci-cubic holographic dark energy in the presence of
interaction with the dark matter sector. Using eqn.(\ref{intm}) we
get the matter energy density as,
\begin{equation}
\rho_{m}=\frac{\rho_{m0}}{a^{3+\eta}}
\end{equation}
Using this the Hubble parameter gets modified as below,
\begin{equation}\label{hubint}
H=\frac{H_{0}\sqrt{\Omega_{m0}}}{\sqrt{a^{3+\eta}\left(1-\Omega_{DE}\right)}}
\end{equation}
The time gradient of the Hubble parameter is consequently modified
as,
\begin{equation}\label{hdotint}
\dot{H}=-\frac{H^{2}}{2\left(1-\Omega_{DE}\right)}\left[\left(3+\eta\right)\left(1-\Omega_{DE}\right)-\Omega_{DE}'\right]
\end{equation}
Further the Ricci and the cubic invariants get respectively
modified as,
\begin{equation}
R=-3H^{2}\left(1-\eta+\frac{\Omega_{DE}'}{1-\Omega_{DE}}\right)
\end{equation}
\begin{equation}
P=3\tilde{\beta}H^{6}\left(\frac{3\Omega_{DE}'}{1-\Omega_{DE}}-\left(5+3\eta\right)\right)
\end{equation}
Using the above invariants the modified form of the differential
equation (\ref{energydiff}) is obtained as,
\begin{equation}
\Omega_{DE}+\frac{3\alpha\left[\left(\eta-1\right)\left(\Omega_{DE}-1\right)+\Omega_{DE}'\right]}
{\Omega_{DE}-1}+3^{1/3}\lambda\left[\frac{\tilde{\beta}\left(\left(5+3\eta\right)
\left(\Omega_{DE}-1\right)+3\Omega_{DE}'\right)}{\Omega_{DE}-1}\right]^{1/3}=0
\end{equation}
This equation does not admit any analytic solution and so we
search for approximations that will allow for a solution. We
expand binomially the third term on the left hand side of the
above equation and consider only the linear terms. The solution is
obtained as,
\begin{equation}
\Omega_{DE}=\frac{e^{3\alpha\left(\varrho+C_{4}\right)}+3^{5/3}\alpha~
e^{\varrho+C_{4}+3\alpha\eta\left(\varrho+C_{4}\right)
+3^{1/3}\beta^{1/3}\left(5+3\eta\right)^{1/3}\lambda\left(\varrho+C_{4}\right)}
\left(\eta-1+\frac{\tilde{\beta}^{1/3}\left(5+3\eta\right)^{1/3}\lambda}{3^{2/3}\alpha}\right)}
{e^{3\alpha\left(\varrho+C_{4}\right)}-3^{2/3}
e^{\varrho+C_{4}+3\alpha\eta\left(\varrho+C_{4}\right)
+3^{1/3}\beta^{1/3}\left(5+3\eta\right)^{1/3}\lambda\left(\varrho+C_{4}\right)}}
\end{equation}
where $\varrho=-\frac{\left(5+3\eta\right)x}{15\alpha+9\alpha
\eta+3^{1/3}\tilde{\beta}^{1/3}\left(5+3\eta\right)^{1/3}\lambda}$
and $C_{4}$ is the constant of integration. For $\eta>0$, we have
a dark energy dominated universe soon enough, but for $\eta<0$
there will be a delayed dark energy dominated epoch.

\subsection{Evolution in the presence of radiation}
In order to properly explain the entire thermal history of the
universe we have to include the radiation component in our set up
and then properly explore the scenario to obtain the radiation
dominated early universe as indicated by the observations. In this
section we will study the evolution equations of the Ricci-cubic
holographic dark energy in presence of radiation. We begin by
considering the density parameter for radiation as,
\begin{equation}
\Omega_{r}\equiv\frac{\kappa^{2}}{3H^{2}}\rho_{r}
\end{equation}
Including the radiation component in the FLRW eqn.(\ref{flrweq1})
we get $\Omega_{m}+\Omega_{DE}+\Omega_{r}=1$. The Hubble parameter
gets consequently modified as,
\begin{equation}\label{hubrad}
H=\frac{H_{0}\sqrt{\Omega_{m0}}}{\sqrt{a^{3}\left(1-\Omega_{DE}-\Omega_{r}\right)}}
\end{equation}
The time gradient of the Hubble parameter is then given by,
\begin{equation}\label{hdotrad}
\dot{H}=-\frac{H^{2}}{2\left(1-\Omega_{DE}-\Omega_{r}\right)}
\left[3\left(1-\Omega_{DE}-\Omega_{r}\right)-\Omega_{DE}'-\Omega_{r}'\right]
\end{equation}
The scalar invariants are modified as follows,
\begin{equation}
R=-3H^{2}\left(1+\frac{\Omega_{DE}'+\Omega_{r}'}{1-\Omega_{DE}-\Omega_{r}}\right)
\end{equation}
\begin{equation}
P=3\tilde{\beta}H^{6}\left(\frac{3\left(\Omega_{DE}'+\Omega_{r}'\right)}
{1-\Omega_{DE}-\Omega_{r}}-5\right)
\end{equation}
Using these results with eqns.(\ref{denrc1}) and (\ref{denpa}) we
get a differential equation in terms of $\Omega_{DE}$ as given
below,
\begin{equation}\label{omega22}
\Omega_{DE}+3^{1/3}\lambda
\left(\frac{\tilde{\beta}\left(-5+5\Omega_{DE}+3\Omega_{DE}'+5\Omega_{r}+3\Omega_{r}'\right)}{\Omega_{DE}+\Omega_{r}-1}\right)^{1/3}
-\frac{3\alpha\left(-1+\Omega_{DE}-\Omega_{DE}'+\Omega_{r}-\Omega_{r}'\right)}{\Omega_{DE}+\Omega_{r}-1}=0
\end{equation}
The equations (\ref{rdashf}) and (\ref{pdashf}) are modified as
follows,
\begin{equation}\label{rdashfrad}
\frac{R'}{3H^2}=3+2\left(\frac{\Omega_{DE}'+\Omega_{r}'}{1-\Omega_{DE}-\Omega_{r}}\right)
-\frac{2(\Omega_{DE}'+\Omega_{r}')^2}{\left(1-\Omega_{DE}-\Omega_{r}\right)^2}
-\frac{\Omega_{DE}''+\Omega_{r}''}{1-\Omega_{DE}-\Omega_{r}}
\end{equation}
and

$$\frac{P'}{9H^{2}P^{2/3}}=\frac{\tilde{\beta}^{1/3}}{3^{2/3}\left(1-\Omega_{DE}-\Omega_{r}\right)^{4/3}
\left[3\left(\Omega_{DE}'+\Omega_{r}'\right)-5\left(1-\Omega_{DE}-\Omega_{r}\right)\right]^{2/3}}
\times\left[\left(\Omega_{DE}''+\Omega_{r}''\right)\left(1-\Omega_{DE}-\Omega_{r}\right)\right.$$

\begin{equation}\label{pdashfrad}
\left.+15\left(1-\Omega_{DE}-\Omega_{r}\right)^{2}+2\left(\Omega_{DE}'+\Omega_{r}'\right)
\left\{2\left(\Omega_{DE}'+\Omega_{r}'\right)+7\left(\Omega_{DE}+\Omega_{r}\right)-7\right\}\right]
\end{equation}
Using the above expressions the EoS parameter given in
eqn.(\ref{eosrgbhde}) gets modified to,
$$w_{DE}=-1-\Omega_{DE}^{-1}\left[\alpha\left\{3+2\left(\frac{\Omega_{DE}'+\Omega_{r}'}{1-\Omega_{DE}-\Omega_{r}}\right)
-\frac{2(\Omega_{DE}'+\Omega_{r}')^2}{\left(1-\Omega_{DE}-\Omega_{r}\right)^2}
-\frac{\Omega_{DE}''+\Omega_{r}''}{1-\Omega_{DE}-\Omega_{r}}\right\}\right.$$

$$\left.-\frac{\lambda\tilde{\beta}^{1/3}}{3^{2/3}\left(1-\Omega_{DE}-\Omega_{r}\right)^{4/3}
\left[3\left(\Omega_{DE}'+\Omega_{r}'\right)-5\left(1-\Omega_{DE}-\Omega_{r}\right)\right]^{2/3}}
\times\left[\left(\Omega_{DE}''+\Omega_{r}''\right)\left(1-\Omega_{DE}-\Omega_{r}\right)\right.\right.$$

\begin{equation}\label{eosrgbhde2}
\left.\left.+15\left(1-\Omega_{DE}-\Omega_{r}\right)^{2}+2\left(\Omega_{DE}'+\Omega_{r}'\right)
\left\{2\left(\Omega_{DE}'+\Omega_{r}'\right)+7\left(\Omega_{DE}+\Omega_{r}\right)-7\right\}\right]\right]
\end{equation}

In this case eqn.(\ref{omega22}) does not admit any analytical
solution and one should elaborate it numerically. In Fig.(6) we
have presented the behaviour of the Ricci-cubic HDE density
parameter $\Omega_{DE}$, the matter density parameter $\Omega_{m}$
and the radiation density parameter $\Omega_{r}$ against the
redshift. In order to comply with the observations we have imposed
the constraints $\Omega_{DE}(z=0)\equiv\Omega_{DE0}\approx 0.68$,
$\Omega_{m}(z=0)\equiv\Omega_{m0}\approx 0.32$ and
$\Omega_{r}(z=0)\equiv\Omega_{r0}\approx 0.0001$ at the present
time. It is seen from the figure that with the evolution of the
universe ($z\rightarrow 0$) filled with the HDE fluid there is a
transition of the universe from a matter dominated to a dark
energy dominated epoch, with the radiation slowly decaying out.
The universe sequentially evolves into radiation dominated,
matter-dominated and finally, a dark energy-dominated regime which
are in agreement with known results.

\begin{figure}\label{fig6}
~~~~~~~~~~~~~~~~~~~~~~~~~~~~~~\includegraphics[height=2.2in,width=3.5in]{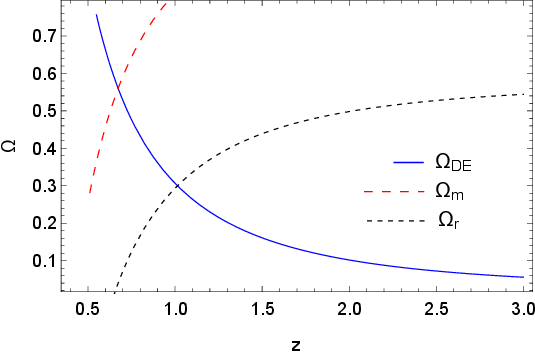}~~~~~\\

~~~~~~~~~~~~~~~~~~~~~~~~~~~~~~~~~~~~~~~~~~~~~~~~~~~~~~~~~~~~Fig.6~~~~~~~~~~~~~~~~~~~~\\

\vspace{1mm} \textit{\textbf{Fig.6} shows the evolution of the
dimensionless density parameters $\Omega_{DE}$, $\Omega_{m}$ and
$\Omega_{r}$ of the various components of the universe as a
function of the redshift $z$. Here we have imposed the conditions
$\Omega_{DE}(z=0)\equiv\Omega_{DE0}\approx 0.68$,
$\Omega_{m}(z=0)\equiv\Omega_{m0}\approx 0.32$ and
$\Omega_{r}(z=0)\equiv\Omega_{r0}\approx 0.0001$ at the present
time. We have taken $\alpha=-0.2, \lambda=0.02,
\tilde{\beta}=0.01$.}
\end{figure}

%%%%%%%%%%%%%%%%%%%%%%%%%%%%%%%%%%%%
\section{Discussion \& Conclusion}
%%%%%%%%%%%%%%%%%%%%%%%%%%%%%%%%%%%%
In this work, we have proposed the Ricci-cubic holographic dark
energy, inspired by the cubic curvature invariant. Here the
infrared cutoff is determined by a combination of the Ricci scalar
and the cubic curvature invariant. The advantage of such a
formulation over the standard ones (like the Ricci holographic
dark energy or the agegraphic dark energy model) is that, here the
infrared cutoff and consequently the holographic dark energy
density is not dependent on the future or the past evolution of
the universe. The complete evolution is dependent on the current
features of the universe, which is big advantage over the standard
formulations. One more advantage is that here the infrared cutoff
is given by invariants, which are of fundamental theoretical
importance in gravity theories. Adopting a generalized approach we
included contributions from both the Ricci scalar and cubic
curvature invariant of the same order in the infrared cutoff and
consequently in the energy density of the holographic dark energy.

First of all we formed the model and set up the necessary
equations of the Ricci-cubic holographic dark energy. Then
considering the Friedmann metric we proceeded to study the
cosmological evolution of the universe filled by the Ricci-cubic
holographic dark energy. Suitable dimensionless density parameters
for matter and dark energy were constructed. A differential
equation involving the dimensionless density parameter of the
holographic dark energy was formed in terms of the logarithm of
the scale factor (directly related to the logarithm of redshift).
Under some assumptions this equation admitted an analytical
solution. We also determined the equation of state parameter and
the deceleration parameter in terms of the dimensionless density
parameters.

The Cosmological features of the Ricci-cubic holographic dark
energy model are quite interesting and presents some increased
capabilities in its dynamics. This is due to the presence of two
model parameters controlling the two curvature invariants. Plots
were generated for the cosmological parameters and it was seen
that they comply with the cosmological evolution of the universe.
It was also seen that the onset of the accelerated expansion can
be fine tuned at $z\approx 0.45$ which is in agreement with the
observations. Various plots for the equation of state parameter
were generated to explore the effects of the model parameters on
the evolution of the universe. It was seen that for suitable
parameter spaces, it is possible to have complete quintessence
like behaviour, complete phantom like behaviour, and
phantom-divide crossing during the cosmological evolution. The
equation of state parameter can also take the value exactly $-1$
mimicking the $\Lambda$CDM universe. An asymptotic de-Sitter like
evolution is also well supported by this model under suitable
conditions. When the contribution from the cubic invariant is set
to zero, we get the usual Ricci holographic dark energy. Moreover
when the contribution from the Ricci scalar is set to zero, we get
the "pure" cubic holographic dark energy. Finally we included
radiation in our set-up, and it was seen that, sequentially
radiation, matter and dark energy epoch are possible. In the end
the universe entered into a dark energy dominated regime,
resulting in the late cosmic acceleration. It should be mentioned
here that each plot is generated by using a different set of
parameter values to comply with the known results. In future this
can be well-adjusted by constraining the parameters using
observational data.

Eventually we conclude by stating that there are some future work
that needs to be done to deeply understand the nature of
Ricci-cubic holographic dark energy, and consequently consider it
as a successful candidate of dark energy. To explore the global
behaviour of the scenario at late times, we need to employ the
dynamical system analysis. In order to constrain the model
parameters we need to use observational data from Type Ia
supernovae (Sn Ia), baryon acoustic oscillations (BAO), cosmic
microwave background (CMB) shift parameter and Hubble parameter
observations. Apart from this, in order to check the model
stability we need to perform a perturbation analysis. These
necessary studies are promising future projects related to
Ricci-cubic holographic dark energy model.

%%%%%%%%%%%%%%%%%%%%%%%%%%%%%
\section*{Data Availability Statement}
%%%%%%%%%%%%%%%%%%%%%%%%%%%%%
No new data were used or generated during the preparation of this
manuscript.

%%%%%%%%%%%%%%%%%%%%%%%%%%
\section*{Acknowledgments}
%%%%%%%%%%%%%%%%%%%%%%%%%%

The author acknowledges the Inter University Centre for Astronomy
and Astrophysics (IUCAA), Pune, India for granting visiting
associateship. We would like to thank the anonymous referees for
carefully reading the paper and we are grateful for their
constructive comments which substantially helped in improving the
quality of the paper.

%%%%%%%%%%%%%%%%%%%%%%%%%%%%%%%%%%%%%%%%%%%%%%%%%%%%%%%%%%%%%%%%%%%%%
%%%%%%%%%%%%%%%%%%%%%%%%%%%%%%%%%%%%%%%%%%%%%%%%%%%%%%%%%%%%%%%%%5

\end{document}